\documentclass[twocolumn, aps, prb, amsmath,superscriptaddress]{revtex4}

\usepackage{graphicx}

\usepackage{fancyhdr}
\begin{document}

\title{Pumped quantum systems: immersion fluids of the future?}

\author{Vikas Anant}
\author{Magnus R\aa{}dmark}
\author{Ayman F.\ Abouraddy}
\affiliation{Research Laboratory of Electronics, Massachusetts Institute of Technology, Cambridge, Massachusetts 02139}
\author{Thomas C.\ Killian}
\affiliation{ Department of Physics and Astronomy, Rice University, Houston, Texas 77251}
\author{Karl K. Berggren}
\email[Electronic mail:]{berggren@mit.edu}
\affiliation{Research Laboratory of Electronics, Massachusetts Institute of Technology, Cambridge, Massachusetts 02139}
\preprint{Conf.\ Paper EP15.2, JVST Manuscript 28303, AIP ID\# 077506JVB}
\begin{abstract}
Quantum optical techniques may yield immersion fluids with high indices of refraction without absorption.  We describe one such technique in which a probe field experiences a large index of refraction with amplification rather than absorption, and examine its practicality for an immersion lithography application. Enhanced index can be observed in a three-level system with a tunable, near-resonant, coherent probe and incoherent pump field that inverts population of the probe transition.  This observation contradicts the common belief that large indices of refraction are impossible without absorption, however it is well in accord with existing electromagnetic theory and practice.   Calculations show that a refractive index $\gg 2$ is possible with practical experimental parameters.  A scheme with an incoherent mixture of pumped and unpumped atoms is also examined, and is seen to have a lower refractive index ($\sim2$) accompanied by neither gain nor loss.
\end{abstract}

\maketitle

\thispagestyle{fancy}

\section{Introduction}
Immersion lithography is currently the technology of choice for integrated circuit fabrication at the 65-nm-gate-length node.\cite{gil,rothschild} Further scaling of this technology will require either a reduced source wavelength or a higher-refractive-index immersion fluid. But efforts to make economical sources and optics at wavelengths shorter than 193 nanometers have so far been unsuccessful.  Additionally, refractive-index scaling is often denegrated based on classical arguments that index enhancement near an atomic resonance is always accompanied by absorption. Thus it may appear that device scaling is nearing its end. However, when atomic resonances are treated quantum-mechanically, the connection between absorption and index enhancement can be circumvented. In this paper, we apply quantum-optical techniques to the problem of index enhancement for continued scaling of optical lithography.

We present a scheme that can yield an enhanced index of refraction ($n\gg1$) with gain instead of absorption.\cite{va-apl-paper}  In this scheme, a near-resonant probe laser experiences a high refractive index in a pumped quantum system.  The refractive index is a function of both the intrinsic properties of the quantum system, such as spontaneous decay rates, and the controllable properties, such as atomic density.\cite{why-atomic}  A few possible applications of this scheme are immersion lithography,\cite{gil,rothschild} microscopy,\cite{microscopy} all-optical switching, and optical buffering\cite{delay}.  While previous proposals for phase-coherent quantum-optics-based index-enhancement techniques promise an appreciable refractive index with no absorption,\cite{eit1,eit2,eit3,eit4,eit5,lukin} only enhancements of $\Delta n \approx 10^{-4}$ have in fact been demonstrated due to practical experimental obstacles.\cite{zibrov}  On the other hand, our calculations show that a maximum refractive-index of $\sim$ 6 is possible with our scheme for reasonable experimental parameters.
In principle, the refractive index can be enhanced beyond 6 with a higher atomic density.  In addition, our scheme should prove easier to implement than existing schemes based on quantum interference because the pump and probe lasers do not need to be collinear or phase-locked.  Thus our proposed scheme is likely to be more practical and more effective than existing quantum-optical schemes at achieving enhanced index without absorption. 

To apply the proposed scheme to immersion lithography, several issues would  have to be resolved.  An immersion medium consisting of a pumped quantum system would generate poor image contrast due to amplified spontaneous emission.  Additionally, due to power broadening, the refractive-index rapidly decays from its enhanced value to unity in a short propagation distance.  These issues are discussed in Sec.\ \ref{sec.hurdles}.  A  solution to some of the problems is suggested in Sec.\ \ref{sec.mixture} where we describe a system composed of a mixture of quantum gases described by \emph{absorptive} two-level quantum systems and \emph{amplifying} (pumped) three-level quantum systems probed by a single laser.\cite{eit3}  The mixture exhibits enhanced index of refraction without loss \emph{or gain}.

\section{Pumped three-level system: high-$n$ with gain}

A simple pumped atomic resonance illustrates the concept of refractive-index enhancement with gain.  In an unpumped system, a probe beam tuned near resonance experiences a high refractive index, but is  absorbed in a fraction of a wavelength.  In our scheme, we use a pump laser to invert population on the probe transition.  A probe beam interacting with a population-inverted transition also experiences a high refractive index but is amplified instead of absorbed.  In the following section, we specify the  parameters of our system.

\subsection{Energy level structure}

We chose to illustrate our index-enhancement scheme in a three-level system as shown in Fig.\ 1, although other three-level  or four-level systems will also suffice.  The three-level system shown in Fig.\ 1 has a ground state $\lvert a \rangle$, excited state $\lvert b \rangle$ and upper lying level $\lvert c \rangle$.   A pump field $\omega_\text{pump}$ is used to invert population on the $\lvert a \rangle$ - $\lvert b \rangle$ transition via an incoherent deexcitation from $\lvert c \rangle$ to $\lvert b \rangle$.    A probe field detuned from the $\lvert b \rangle$ - $\lvert a \rangle$ transition by frequency $\Delta$ will see an enhanced index.  In the following section, we quantify the amount of refractive index enhancement and show that the probe field will also experience amplification rather than absorption due to population inversion.

\begin{figure}
	\centering
	\includegraphics{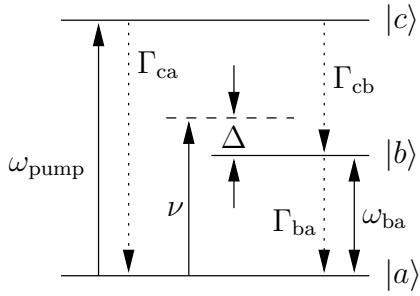}
	\caption{Energy-level diagram for a pumped three-level system showing driving fields (probe $\nu$ and pump $\omega_\text{pump}$) and incoherent decay rates $\Gamma_{ij}$ corresponding to decay from level $\lvert i \rangle$ to $\lvert j \rangle$.  The probe field is detuned by $\Delta$ from the $\lvert b \rangle$ - $\lvert a \rangle$ resonance.}
	\label{fig.energy-levels}
\end{figure}

\subsection{Index enhancement for unbroadened thin media}\label{sec.pumped-calc}

\chead{}

To calculate the refractive index as seen by the probe laser, we formulate the density matrix equations of motion\cite{scully-qo} for the system.  The population in level $\lvert i \rangle$ is denoted by the on-diagonal density matrix element $\rho_{ii}$ and decays into another level $\lvert j \rangle$ at rate $\Gamma_{ij}$.  The coherence between levels $\lvert i \rangle$ and  $\lvert j \rangle$ is given by $\rho_{ij}$ and decays at rate $\gamma_{ij}$.  A phenomenological treatment of spontaneous decay rates and coherence decay rates leads to the following equations:

\begin{subequations}
	\label{eq.dme}
	\begin{eqnarray}
		\label{eq.dmei}
		\dot{\rho}_\text{aa} &=&
		(\Gamma_\text{ca}-\Gamma_\text{ba})\rho_\text{cc} - (r_\text{ac} + \Gamma_\text{ba})\rho_\text{aa} 
		\\&&\nonumber + \frac{i\Omega_\text{R}}{2}(\rho_\text{ba}-\rho_\text{ab}) +\Gamma_\text{ba}
		,\\%
		\dot{\rho}_\text{ab} &=&
		-\frac{i\Omega_\text{R}}{2}(\rho_\text{cc}+2\rho_\text{aa}-1)-(\gamma_\text{ab}+i\Delta)\rho_\text{ab}
		,\\%
		\dot{\rho}_\text{ca} &=&
		-\bigl( \gamma_\text{ca} + i (\omega_\text{cb}-\Delta)\bigr)\rho_\text{ca} - \frac{i\Omega_\text{R}}{2}\rho_\text{cb}
		,\\%
		\dot{\rho}_\text{cb} &=&
		-\frac{i\Omega_\text{R}}{2}\rho_\text{ca}- (\gamma_\text{cb} + i \omega_\text{cb})\rho_\text{cb}
		,\\%
		\dot{\rho}_\text{cc} &=& -(\Gamma_\text{ca}+\Gamma_\text{cb})\rho_\text{cc} + r_\text{ac}\rho_\text{aa}
		,%
		\label{eq.dmef}
	\end{eqnarray}
\end{subequations}
where $r_\text{ac}$ is the pump rate, $\Omega_\text{R}$ is the Rabi frequency of the $\lvert b \rangle$ - $\lvert a \rangle$ transition, and we have used $\rho_\text{aa}+\rho_\text{bb}+\rho_\text{cc} = 1$.  

The susceptibility $\chi$ as seen by the probe is related to the coherence between states $\lvert b \rangle$ and $\lvert a \rangle$ given by the off-diagonal density matrix element $\rho_\text{ba}$ (see Ref.\ \onlinecite{scully-qo} for more details).  Equations \eqref{eq.dmei}-\eqref{eq.dmef} were solved under the rotating-wave approximation and steady-state conditions to yield the following expression for $\chi$:
	\begin{equation}
				\chi = N \Gamma_{\text{ba}}  \frac{3}{8 \pi^2} \lambda_{\text{ba}}^3 
				\frac{1}
				{\Delta+i \gamma_\text{ab}}
				(\rho_{\text{bb}}-\rho_{\text{aa}}),
								\label{eq.chi}
	\end{equation}
	where the population difference between states $\lvert b \rangle$ and $\lvert a \rangle$ is given by
	\begin{equation}
 				\rho_{\text{bb}}-\rho_{\text{aa}} = \frac{
 				\Gamma_\text{cb} r_\text{ac} - \Gamma_\text{ba} \Gamma_\text{c}
 				}{
 				r_\text{ac} \Gamma_\text{cb} + \Gamma_\text{ba}(\Gamma_\text{c}+r_\text{ac}) + \eta (r_\text{ac} + 2 \Gamma_\text{c})
 				}.
								\label{eq.popdiff}
	\end{equation}
Here $\eta = \Omega_\text{R}^2 \gamma_\text{ab}/2(\Delta^2 + \gamma_\text{ab}^2)$ and $\Gamma_\text{c} = \Gamma_\text{cb}+\Gamma_\text{ca}$.  From Eqs.\ \eqref{eq.chi} and \eqref{eq.popdiff}, we see that $\chi$ depends on controllable parameters, namely $N$, $\Delta$, $\Omega_\text{R}$ and $r_\text{ac}$, and parameters that are intrinsic to the atomic species, namely decay rates and the probe transition wavelength $\lambda_\text{ba}$.  In order to illustrate the magnitude of refractive index enhancement, we use a three level system with $\Gamma_\text{ba} = 2.2 \times 10^8 \text{s}^{-1}$ and $\lambda_\text{ba} =$ 422.8 nm.\cite{atomic-ca-revised}  In our calculations, the $\lvert c \rangle - \lvert a \rangle$ and $\lvert c \rangle - \lvert b \rangle$ transitions were treated phenomenologically with parameters that enabled a large population inversion.

Figure 2 shows the frequency dependence of the susceptibility $\chi$ for an unpumped and pumped system with $\chi$ taken to first order in $\Omega_\text{R}$.  For these plots, $\chi = \chi^\prime + i \chi^{\prime\prime}$, where $\chi^{\prime\prime}>0$ implies an absorptive medium, while $\chi^{\prime\prime}<0$ implies an amplifying medium.  Propagation effects were not taken into account, so these plots apply only for infinitesimally thin media.  Figure 2(a) was obtained by expanding Eq.\ \eqref{eq.popdiff} to first order in $\Omega_\text{R}$, setting $r_\text{ac}=0$, and substituting the result into Eq.\ \eqref{eq.chi}.   From this plot, it is evident that a large susceptibility is possible, but at the cost of high absorption.  Figure 2(b) was obtained via an identical analysis except that a value for $r_\text{ac}$ was chosen to ensure population inversion between levels $\lvert b \rangle$ and $\lvert a \rangle$.   $\chi^{\prime\prime}$ shown in Fig.\ 2(b) has the opposite sign of the $\chi^{\prime\prime}$ curve shown in Fig.\ 2(a) but $\chi^\prime$ is changed only by a reflection in the detuning axis, signifying that a very high susceptibility can be obtained with \emph{gain} rather than \emph{absorption}.  This is one of the key points of this paper. 

\begin{figure}
	\centering
	\includegraphics{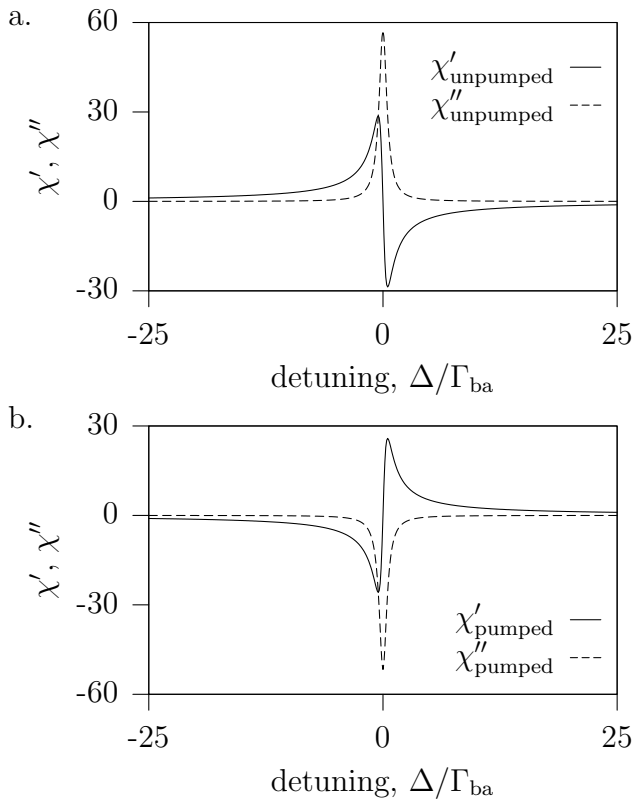}
	\caption{Plot of the susceptibility $\chi = \chi^\prime + i \chi^{\prime\prime}$ calculated to first order in $\Omega_\text{R}$ for (a)  an unpumped system, and (b) a pumped system with population inversion. Parameters are $N = 10^{16}\ \text{cm}^{-3}, 
		\gamma_{\text{ab}}=0.5 \Gamma_{\text{ba}},
		\Gamma_{\text{ca}}=0.1 \Gamma_{\text{ba}},
		\Gamma_{\text{cb}}=10 \Gamma_{\text{ba}}$, and  $r_{\text{ac}}=200\Gamma_{\text{ba}}$.  We see that $\chi^{\prime\prime}<0$ for the pumped case, resulting in amplification (rather than absorption) of the probe laser.}
	\label{fig.chi-vs-detuning}
\end{figure}

The refractive index can be calculated from the susceptibilities via $n = n^\prime + i n^{\prime\prime} = \sqrt{1+\chi}$ where $n^\prime$ denotes the refractive index, while $n^{\prime\prime}$ is the absorption coefficient of the medium.  For the parameters used for Fig.\ 2(b), at the optimum detuning, the refractive index can be as high as $\sim6$.  It is possible to enhance the refractive index still further: $n^\prime$ can be increased by opting for an atomic species with different intrinsic parameters, e.g. a larger dipole matrix element for the $\lvert b \rangle$ - $\lvert a \rangle$ transition.  Another way to increase $n^\prime$ is by changing the controllable atomic parameters, e.g. by using a vapor with a high atomic density.  The effect of this change is shown in Fig.\ 3, where we have plotted the maximum achievable refractive index $n^\prime_\text{max}$ with respect to detuning as a function of the atomic density.  
As the plot shows, an increase in the atomic density of one order of magnitude from $10^{16}$ to $10^{17}$ atoms/cm$^3$ increases $n^\prime_\text{max}$ by a factor of $\sim$ 3.  The importance of the ability to increase the refractive index with a controllable parameter will become evident in the following section, where we will find that the refractive index drops in response to temperature-related broadening effects.  

\begin{figure}
	\centering
	\includegraphics{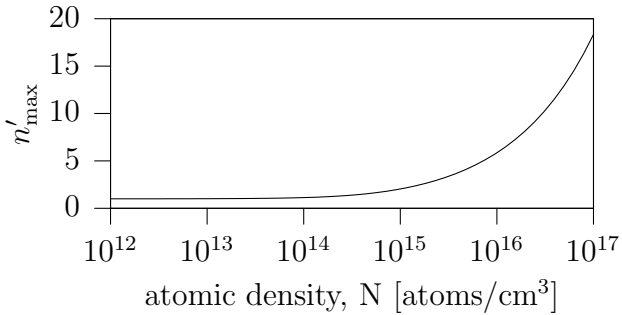}
	\caption{Plot of maximum value of refractive index $n^\prime_\text{max}$ (with respect to detuning) as a function of atomic density $N$.  See Fig.\ 2 caption for other parameters.}
	\label{fig.max-n-vs-N}
\end{figure}

\subsection{Index enhancement for broadened, thin media}

To evaluate the effect of temperature on the refractive index, we consider the effects of collisional, Doppler, and power broadening.  Throughout this subsection, however, we neglect effects of the propagation of the probe beam.  Thus these results apply only for infinitesimally thin media.  Propagation effects will be discussed later.  We will now elaborate on the various broadening mechanisms:
\begin{enumerate}
\item \emph{Non-radiative collisional broadening} is attributed to the random change in the phase of atomic wavefunctions that results from elastic collisions.  Non-radiative collisional broadening dominates over radiative collisional broadening that occurs due to inelastic collisions.\cite{loudon}  Collisional broadening was taken into account by adding an extra decay term in the equations of motion for the off-diagonal matrix element $\rho_{ba}$.  The decay term is given by \cite{loudon}
 $4 d_\text{ctrs}^2 N (\pi k_\text{B} T/M)^{1/2}$,
where $d_\text{ctrs}$ is the average distance between atoms during a collision, $T$ is the temperature, $k_\text{B}$ is the Boltzmann constant, and $M$ is the mass of the colliding atoms.  
\item \emph{Power broadening} occurs for probe intensities so large that the Rabi frequency $\Omega_\text{R}$ is greater than the natural decay rate for the $\lvert b \rangle$ to $\lvert a \rangle$ transition.  Power broadening was taken into account by simply using the full expression in Eq.\ \eqref{eq.popdiff} rather than taking it only to first order in $\Omega_\text{R}$.
\item \emph{Doppler broadening} occurs when a moving atom interacts with the probe field and sees a frequency shift due to the Doppler effect.  Interaction of the probe field with the temperature-dependent velocity distribution of the atomic vapor then leads to broadening of the absorption lineshape.  Doppler broadening was taken into account by numerically performing a convolution integral between the lineshape function for Doppler broadening and the lineshape for susceptibility. The lineshape function for Doppler broadening, when expressed in terms of $\Delta$, is a normal distribution with variance $(2\pi/\lambda_\text{ba})^2 k_\text{B} T/M$.
\end{enumerate}

These three mechanisms together act to broaden and reduce the peak value of the refractive index.  Figure 4 shows a plot of the refractive index as a function of detuning for a pumped system at various temperatures.   The maximum value of the refractive index gets reduced from $\sim6$ at 0K to $\sim1.5$ at room temperature, while the entire curve broadens as the temperature is increased.\cite{footnote-increase-n-N}  While Doppler broadening was the dominant mechanism that lead to a decline in refractive index for the parameters chosen, power broadening contributed more significantly when propagation effects were included.   This issue is discussed in the following section along with the problem of amplified spontaneous emission.

\begin{figure}
	\centering
	\includegraphics{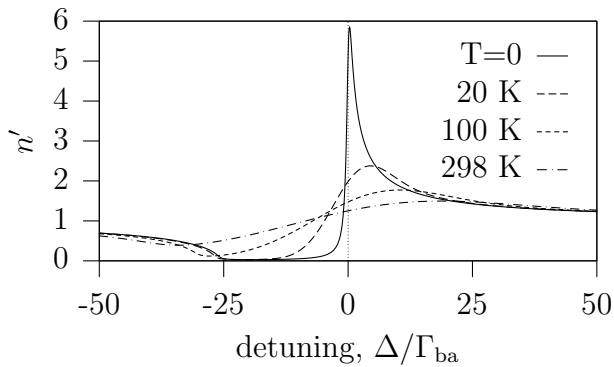}%
	\caption{Plot of refractive-index $n^\prime$ as a function of detuning for a broadened system for various temperatures.  The numerical calculation incorporated collisional, power, and Doppler broadening mechanisms for an infinitesimally small propagation distance of the probe beam.  Parameters are $d_\text{ctrs} = 3.6 \text{\AA}$, probe laser intensity $I_0 = 1 ~\text{mW/cm}^2$, $M=40~\text{a.m.u}$.  See Fig.\ 2 caption for other parameters.}
	\label{fig.n-vs-detuning-broadening}
\end{figure}

\section{Implementation hurdles}\label{sec.hurdles}

The engineering issues that need to be addressed in order for a pumped atomic system to be used as an immersion medium are:
\begin{enumerate}
\item the medium has to be chemically and physically compatible with conventional immersion-lithography systems and processing techniques;
\item the pumping scheme geometry must not inadvertantly expose the photoresist.  A pump field directed towards the photoresist may be a problem if the pump frequency falls within the exposure spectrum of the photoresist;
\item it may not be practical to have an immersion medium with refractive indices higher than $\sim2$  due to the current unavailability of high-index photoresists.  One solution may be to use thin resists that can be exposed using evanescent fields.\cite{alkaisi01}  This problem will be faced by any index-enhancement scheme used for immersion lithography.
\end{enumerate}

In this paper, we will not discuss these issues in detail because we believe that with appropriate chemical and optical engineering, they could in principle be resolved.  Instead we will focus for now on two potential `show-stoppers,' both a consequence of  high gain in the medium.  In the following section, we show that power broadening reduces the initially high refractive index within a very short propagation distance.  In Sec.\ \ref{sec.ase}, we discuss the effect of amplified spontaneous emission in an immersion lithography application.

\subsection{Power broadening}

While power broadening had a negligible effect for an infinitesimally thin medium, it becomes the dominant mechanism that lowers the refractive index when propagation of the probe field in a macroscopic immersion medium is considered.  A qualitative understanding comes from the following reasoning: as the intensity of the probe beam increases (i.e. $\Omega_\text{R}$ increases) due to amplification, the population difference (see Eq.\ \eqref{eq.popdiff}), and thus the refractive index, both decrease.   The effect can be calculated for a propagation distance $z$ within a sample by numerically solving the differential equation 
\begin{align}
	\frac{dI}{dz} &= -\frac{4\pi n^{\prime\prime}I}{\lambda_\text{ba}}, \label{eq.didz}
\end{align}
where $I$ is the intensity of the probe beam.  Note that $n^\prime$ and $n^{\prime\prime}$ are both functions of $I$ and therefore $z$.  The result of Runge-Kutta integration of Eq.\ \eqref{eq.didz} for various detunings is plotted in Fig.\ 5.  The plot reveals an inherent trade-off between refractive index and propagation distance: if one is willing to settle for a lower refractive index (achieved by increasing the detuning $\Delta$), then the maximum distance in the medium in which that index can be maintained (before $n^\prime$ decays rapidly to 1) is increased.  However, for technologically interesting indices of refraction (greater than 2), the maximum thickness is $\sim4 ~\mu \text{m}$ for an atomic density of $10^{16}\,\text{atoms}/\text{cm}^3$.  A medium only 4 $\mu$m thick may present arduous practical challenges to implement in an immersion lithography tool.   The following section explores another critical challenge to implementation.

\begin{figure}
	\centering
	\includegraphics{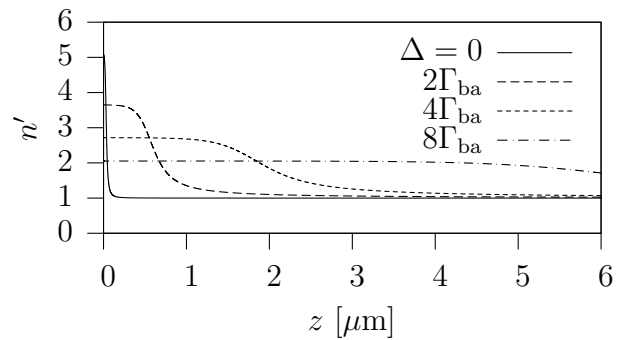}
	\caption{Refractive index $n^\prime$ as a function of propagation distance $z$ in the atomic medium for various detunings $\Delta$.  The index of refraction decreases as a function of $z$ due to an increase in power broadening as the propagating probe beam undergoes amplification.  Parameters used are $I[z=0] =10^3\, \text{W}/\text{m}^2$ and $N=10^{16}\, \text{atoms/cm}^{3}$.   See Fig.\  2 caption for other parameters.}
	\label{fig.n-propagation}
\end{figure}

\subsection{Amplified spontaneous emission}\label{sec.ase}

Amplified spontaneous emission is a problem endemic to many optical systems, including Erbium-doped fiber amplifiers (EDFA),\cite{bjarklev} and master oscillator power amplifiers (MOPA).\cite{mopa}  In our system, the gain medium amplifies photons not only in the probe beam, but also photons that originate from spontaneous-emission events.  In one simulation, we found that spontaneous emission was amplified by a factor of $\sim 100$ in a propagation distance of only $1\,\mu \text{m}$.  This magnitude of amplification is undesirable for two reasons:  (1) an atom that has undergone stimulated emission from a spontaneous emission photon is unavailable for emission stimulated by a probe-beam photon, thus the population inversion is reduced, leading to a reduction in the refractive index of the medium;  and (2) a large amount of noise from spontaneous emission reduces the image contrast in an immersion lithography application.  In the following section, we discuss a scheme that mitigates both this effect and power-broadening issues.
	
\section{A scheme for high-$n$ with neither gain nor loss}\label{sec.mixture}

One solution to the power broadening and amplified spontaneous emission problems is to find a scheme that eliminates gain altogether, yet still retains enhanced refractive index.  Electromagnetically induced transparency would achieve this,\cite{eit1} as would a simple mixture of absorptive and amplifying high-index components in the medium.\cite{eit3}   We examine such a scheme in the following section, using the results of Fig.\ 2, which shows  that absorptive unpumped systems and amplifying pumped systems both exhibit high refractive indices near resonance.

\subsection{Energy level structure}

We examine a scheme with a homogeneously-distributed, equal mixture of absorptive two-level systems and amplifying three-level systems probed concurrently by a single probe beam.  Fig.\ 6 shows the energy level structure of such a scheme.  The mixture consists of uncoupled two- and three-level systems with slightly different resonant frequencies on the probe transition ($\omega_\text{ba} \neq \omega_\text{ed}$).  The two-level system is detuned from the probe field by $\Delta_\text{u}$ while the three-level system is detuned by $\Delta_\text{p}$.  We now describe the calculation of refractive index for this system.

\begin{figure}
	\centering
	\includegraphics{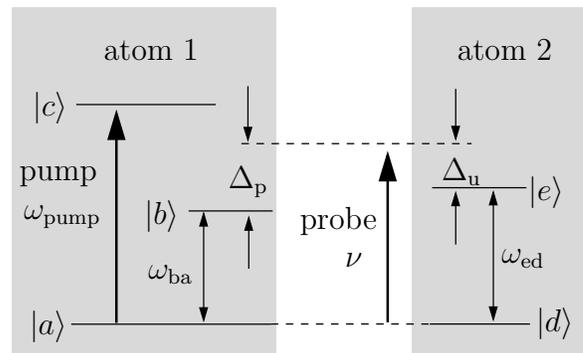}
	\caption{Energy-level diagram a of \emph{pumped} three-level (atom 1) and \emph{unpumped} two-level (atom 2) quantum system.   In the intended mixture, both three- and two-level systems are probed concurrently at frequency $\nu$ and are assumed to be uncoupled and homogeneously distributed within the probe interaction region.}
	\label{fig.mixture-energy-levels}
\end{figure}

\begin{figure}
	\centering
	\includegraphics{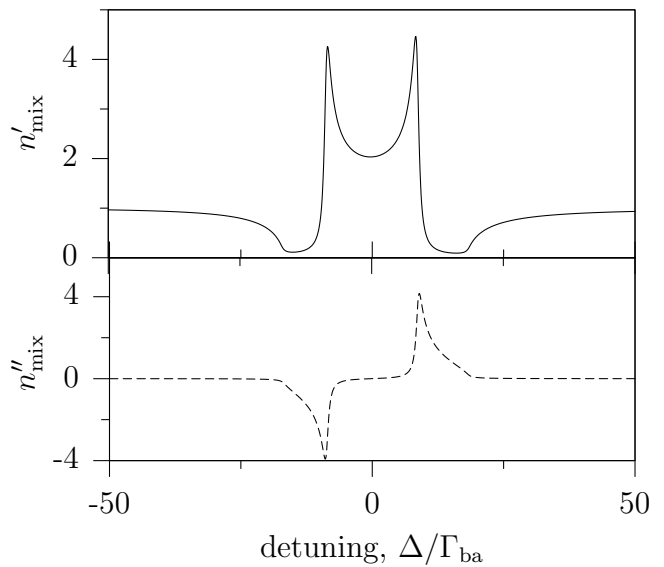}
	\caption{Plot of refractive index $n^\prime_\text{mix}$ and absorption coefficient $n^{\prime\prime}_\text{mix}$ as a function of detuning for a homogeneous equal mixture of uncoupled two- and three-level systems as pictured in Fig.\ 6.  Note that $\Delta = (\Delta_\text{u}+\Delta_\text{p})/2$ and that $n^\prime_\text{mix}$ and $n^{\prime\prime}_\text{mix}$ were calculated to first order in $\Omega_\text{R}$.  Other parameters are $N_\text{p}=N_\text{u}=0.5 \times 10^{16}$ atoms/cm$^3$, $\Gamma_\text{ba}=\Gamma_\text{ed}$, and $\omega_\text{ed}-\omega_\text{ba}=52.7$ GHz.  See \mbox{Fig.\ 2} caption for remaining parameters.  $n^{\prime\prime}_\text{mix}=0$ implies a medium transparent to the probe, which passes through it without gain or loss.}
	\label{fig.n-mixture}
\end{figure}

\subsection{Calculation details and result}

The calculation for the refractive index for this scheme is a simple extension of the calculation for the pumped system given in Sec.\ \ref{sec.pumped-calc}.   We first found the susceptibilities for both pumped and unpumped systems separately, denoted by $\chi_\text{p}$ and $\chi_\text{u}$ respectively.  To find $\chi_\text{p}$ as a function of $\Delta_\text{p}$, we substituted parameters specific to the pumped system (e.g.\ detuning $\Delta_\text{p}$, wavelength of probe transition $\lambda_\text{p}$, atomic density $N_\text{p}$) in Eqs.\ \eqref{eq.chi} and \eqref{eq.popdiff}.  The susceptibility for the unpumped system was found by letting $r_\text{ac}=0$ in Eq.\ \eqref{eq.popdiff} and substituting parameters specific to the unpumped system ($\Delta_\text{u}, \lambda_\text{u}, N_\text{u}$).  We then found the susceptibility of the mixture $\chi_\text{mix} = \chi_\text{p}+\chi_\text{u}$, and expressed $\chi_\text{mix}$ in terms of the average detuning $\Delta = (\Delta_\text{u}+\Delta_\text{p})/2$.  Finally, we computed the refractive index as seen by the probe field from $n_\text{mix} = \sqrt{1+\chi_\text{mix}}$.

Figure 7 shows the dependence of refractive index on detuning for a system with an equal mixture of pumped and unpumped atoms.  The plot of $n^{\prime\prime}_\text{mix}$ shows that there is a region of absorption where $n^{\prime\prime}_\text{mix}>0$ and a region of gain where $n^{\prime\prime}_\text{mix}<0$.  Corresponding to the minima and maxima of the $n^{\prime\prime}_\text{mix}$ curve are resonant peaks in the plot of $n^\prime_\text{mix}$: one corresponding to the pumped species and the other to the unpumped atomic species.  The most interesting and important feature of this plot is that there is a point on the $n^{\prime\prime}_\text{mix}$ plot where $n^{\prime\prime}_\text{mix}=0$.  For that detuning, a probe field propagating in the medium would experience neither gain nor loss.  In effect, the medium becomes \emph{transparent} to the probe field but by a very different mechanism than electromagnetically induced transparency.\cite{eit3} Moreover, at the transparency point the refractive index is greatly enhanced: for this example it is more than 2.

\section{Conclusion \& future work}

The central result of this paper is the presentation of a scheme for refractive-index enhancement ($n^\prime\sim6$ for the example system) accompanied by gain rather than absorption.  We showed that one can achieve extremely high refractive indices, but only for a very thin layer of the atomic medium.  We identified the main implementation hurdles as power broadening and amplified spontaneous emission, both due to the high gain experienced by the probe beam.  Finally, we examined a modified scheme consisting of a mixture of pumped and unpumped systems to tackle these problems and to achieve enhanced refractive index without either gain or loss.

We believe that quantum-optical techniques that take advantage of high susceptibilities near atomic resonances may not only result in a high-index medium for next-generation immersion lithography, but also represent a new direction for refractive-index engineering.  Indeed, an all-optically-controllable refractive index can be useful for applications in optical communications and microscopy.  Future work will concentrate on finding quantum systems with which we can experimentally verify the predictions made in this paper.

The authors gratefully acknowledge partial support by AFOSR.  The authors thank Dr.\ Ying-Cheng Chen  for useful discussions.

\bibliography{077506jvb}

\begin{thebibliography}{20}
\bibitem[{ato()}]{atomic-ca-revised}
\bibinfo{note}{The values for $\Gamma_\text{ba}$ and $\lambda_{ba}$ are
  representative of typical atomic parameters. They correspond to the
  $4^1\text{S}_0$ ($\lvert a \rangle$) to $4^1\text{P}_1$ ($\lvert b \rangle$)
  transition in atomic calcium. Level $\lvert c \rangle$ is assumed to be an
  upper lying energy level which can decay to $\lvert b \rangle$ via multiple
  decays. The analysis is not substantially different for other atomic species
  that can be modelled by a three-level system.}
\bibitem[{foo()}]{footnote-increase-n-N}
\bibinfo{note}{As discussed previously, we may enhanced the refractive index
  further by an increase in atomic density.}
\end{thebibliography}

\end{document}